\documentclass[pra,aps,twocolumn,showpacs]{revtex4}

\usepackage{bm}
\usepackage{amssymb}
\usepackage{graphicx}

\begin{document}

\title{Experimental realization of programmable quantum gate}

\author{Michal Mi\v{c}uda}

\author{Miroslav Je\v{z}ek}

\author{Miloslav Du\v{s}ek}

\author{Jarom\'{i}r Fiur\'{a}\v{s}ek}

\affiliation{Department of Optics, Palack\'{y}
University, 17. listopadu 50, 77200 Olomouc, Czech Republic}


\begin{abstract}
We experimentally demonstrate a programmable single-qubit quantum gate. This device 
applies a unitary phase shift operation to a data qubit with the value of the phase shift 
being fully determined by the state of a program qubit.
Our linear optical implementation is based on the encoding of qubits into polarization states of single photons, two-photon interference on a polarizing beam splitter, and measurement on the output program qubit. We fully characterize the programmable gate by quantum process tomography. The achieved average quantum process 
fidelity exceeding $97\%$ illustrates very good performance of the gate for all values of the encoded phase shift. We also show that by using a different set of program states the device can operate as a programmable partial polarization filter.

\end{abstract}

\pacs{03.67.-a, 42.50.Ex}

\maketitle

Classical computers rely on the combination of a fixed hardware and a flexible
software. The operation performed on the data register is fully determined by
the information stored in the program register and can be altered at will by the user. 
It is intriguing  to attempt to generalize this concept to quantum computing \cite{Nielsen00}.
Imagine a fixed universal quantum processing unit where the transformation of the data qubits
is specified by the quantum state of program qubits. In a seminal paper \cite{Nielsen97} 
Nielsen and Chuang proved that an $n$ qubit quantum register can perfectly encode at most
$2^{n}$ distinct quantum operations. Since already
unitary transformations on a single qubit form the $\mathrm{SU}(2)$ group with uncountably many
elements, this bound seems to severely limit the universality of programmable quantum
gates.

Although perfect universal programmability is ruled out, 
it is nevertheless  possible to construct approximate programmable quantum gates 
and optimize their performance for a given size of the program register
  \cite{Vidal02,Hillery02a,Hillery02b,Hillery04,Brazier05,Hillery06}. 
Two complementary approaches to this problem were pursued in the literature. 
One option is to design gates that operate deterministically,
i.e. always provide an output, but add some noise to the output state \cite{Hillery06}. 
An alternative strategy avoids the extra noise at a cost of reduced success probability    \cite{Nielsen97,Vidal02,Hillery02a}. The gate  then involves a
 measurement whose outcome heralds its success or failure. If restricted to 
successful cases, the gate operates perfectly and noiselessly.

A very important elementary programmable quantum gate was proposed by Vidal, Masanes, and
Cirac (VMC) \cite{Vidal02}. They considered programmable rotation of a single qubit along the $z$ axis of the Bloch sphere,
\begin{equation}
U(\phi) =|0\rangle \langle 0|+e^{i\phi}|1\rangle \langle 1|.
\label{Uphi}
\end{equation}
Here  $|0\rangle$ and $|1\rangle$ denote the computational basis states of the qubit.
In the simplest version of VMC protocol, the phase shift $\phi$ is encoded into a
state of single-qubit program register, 
\begin{equation}
|\phi\rangle_P=\frac{1}{\sqrt{2}}(|0\rangle+e^{i\phi}|1\rangle).
\label{Phiinput}
\end{equation}
A C-NOT gate is applied to the 
data qubit  $|\psi\rangle_{D}=\alpha|0\rangle+\beta|1\rangle$ and the program qubit
$|\phi\rangle_P$. This is  followed by measurement on the program qubit in the computational 
basis.
The measurement outcome $|0\rangle$ indicates successful
application of $U(\phi)$ onto the data qubit while the outcome $|1\rangle$ means that
the operation $U(-\phi)$ has been applied. This scheme thus exhibits a success probability of $50\%$
which is the maximum possible with a single qubit program register. By adding further program qubits, the probability of success can be made arbitrary close to unity.  
Note that an exact specification of the phase shift $\phi$ would require infinitely many classical bits. A striking feature of the programmable quantum gate is that the information on $\phi$
is faithfully encoded into  \emph{a single} quantum bit.

While the theory of programmable quantum gates is well established, little attention has been paid to their experimental realizations. The single-qubit  programmable quantum measurement devices \cite{Dusek02,Fiurasek02,Fiurasek03,DAriano04epl,Barnett03,Bergou05,DAriano05,Bergou06,Zhang06,He07}, where  the state of  program qubit
determines the measurement on data qubit, were implemented for single-photons  \cite{Soubusta04} and for nuclear spins in an NMR experiment \cite{Gopinath05}. Also programmable discriminator of coherent states has been reported \cite{Sedlak07,Bartuskova08}. However, to our knowledge, there is no demonstration of a programmable unitary quantum gate for photonic qubits. In this
paper we close this gap between theory and experiment. Specifically, we implement the 
elementary single-qubit programmable gate proposed by VMC \cite{Vidal02}.

Our optical implementation is based on the encoding of qubits into polarization states 
of single photons. We exploit two-photon interference on a polarizing beam splitter (PBS).
Consider the input states of  data and program photons,
\begin{equation}
|\psi\rangle_{D}=\alpha|H\rangle+\beta|V\rangle, \qquad
|\phi\rangle_{P}=\frac{1}{\sqrt{2}}(|H\rangle+e^{i\phi}|V\rangle),
\end{equation}
where $|H\rangle$ and $|V\rangle$ denote the horizontal and vertical linear polarization states, respectively. Suppose that the PBS  totally transmits horizontally polarized
photons and reflects vertically polarized photons. If we restrict ourselves to the cases
when a single photon emerges in each output port of the PBS \cite{Pittman01,Pittman02,Pittman03}, then the conditional two-photon output state reads
\begin{equation}
\frac{1}{\sqrt{2}}(\alpha|H\rangle_D|H\rangle_P+\beta e^{i\phi}|V\rangle_D |V\rangle_P).
\end{equation}
If the program qubit is measured in the diagonal linear polarization basis 
$|\pm\rangle=\frac{1}{\sqrt{2}}(|H\rangle\pm|V\rangle)$ then the 
data qubit is prepared according to the measurement result in the state  
\begin{equation}
|\psi_{\mathrm{out}}\rangle_D=\alpha|H\rangle\pm\beta e^{i\phi}|V\rangle.
\label{psiout}
\end{equation}
If the measurement outcome is $|+\rangle$ then the unitary transformation $U(\phi)$ has been applied to 
the data photon.
If the outcome is $|-\rangle$ then the state acquires an extra relative $\pi$ phase shift in the $H/V$ basis. This can be compensated
by a fast electrooptical modulator that applies a relative  phase shift $0$ or $\pi$ 
depending on the measurement outcome \cite{Sciarrino06,Prevedel07}. With the active feed-forward the scheme achieves the success probability 
$50\%$.  This  saturates the bound on the achievable success probability \cite{Vidal02} 
so this linear optical scheme is  optimal. In the experiment we implemented passive version of the scheme without feed-forward. We post-select only the events when the program qubit is projected onto state $|+\rangle$ which reduces the theoretical success probability of the protocol to $25\%$.

\begin{figure}[!t!]
\centerline{\includegraphics[width=\linewidth]{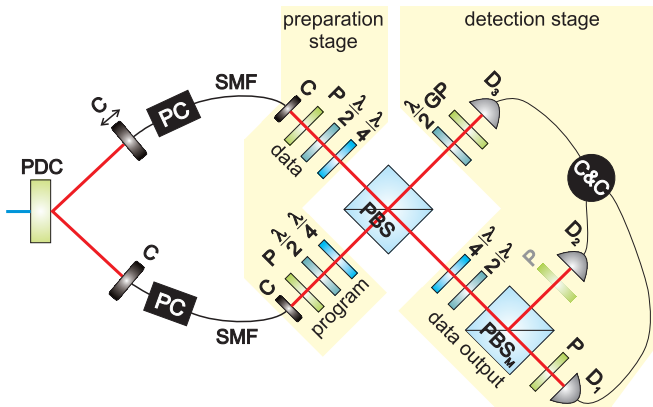}}
\caption{(Color online) Scheme of the experimental setup. The correlated photons
generated in the process of spontaneous parametric down-conversion
(PDC) serve as the program and data qubit. After being prepared in proper
polarization states the photons interfere on the polarizing beam
splitter (PBS). The detection stage consists of polarization analysis,
single-photon detectors (D), coincidence logic and counting module
(C\&C). For polarization setting and analysis the fiber polarization
controllers (PC), half-wave plates ($\frac\lambda2$), quarter-wave plates
($\frac\lambda4$), linear film polarizers (P), Glan polarizer (GP), 
and polarizing beam splitter (PBS$_{\rm M}$) are used. See text for details.}
\label{setup}
\end{figure}

The experimental setup is shown in Fig.~\ref{setup}.
The initial pump beam from continuous-wave laser (CUBE~405~C, Coherent)
with the wavelength of $407\,\rm{nm}$ is focused to $6\,\rm{mm}$ thick
non-linear crystal LiIO$_3$ cut for type I degenerate spontaneous parametric
down-conversion (PDC). After filtering out the
scattered pump light by cut-off filters the down-converted photon
pairs with wavelength of $814\,\rm{nm}$ are coupled (C) into single-mode
optical fibers (SMF) acting as spatial filters. The photons are prepared in horizontal
polarization state at the fibers outputs by fiber polarization
controllers (PC). To achieve maximal polarization purity two
linear film polarizers (P) are employed. 
The required polarization states of both
the data and program photons are set in the preparation stage
by properly rotated wave plates ($\frac\lambda2$, $\frac\lambda4$).
The data and program photons interfere on the polarizing beam
splitter (PBS, Ekspla) and enter the detection stage.
The data photon passes through a sequence of a quarter-wave plate, a half-wave plate,
and another polarizing beam splitter PBS$_\mathrm{M}$ which allows to measure the
photon in an arbitrary polarization basis. The program photon is projected onto the
diagonal linearly polarized state $|+\rangle$ by a half-wave plate and a calcite
Glan polarizer (GP). At the outputs the photons are detected by fiber-coupled
single-photon avalanche photodiodes (SPCM-AQ4C, Perkin Elmer). Detection events
registered by the detectors D$_1$, D$_2$, and D$_3$ are processed by coincidence
logic (TAC/SCA, Ortec) and fed into a counting module (Ortec).

To verify the overlap of photons' spatial mode functions on the PBS we measure
the visibility of Hong-Ou-Mandel (HOM) dip \cite{HOM} in the data output port.
The data and program photons are prepared in horizontal and  vertical polarization
states, respectively, so that they both propagate to the data output.
The wave plates in the data detection stage are set to transform the H/V linear
polarizations onto the diagonal ones. The HOM dip in coincidences between clicks
of detectors D$_1$ and D$_2$ is measured as a function of the time delay between
the program and data photons introduced by a motorized translation of one of the
fiber coupling systems. The measured HOM dip visibility is above $89\%$,
limited mainly by imperfections of PBS$_\mathrm{M}$. Particularly, if we insert
two linear film polarizers in front of APDs $D_1$ and $D_2$ the visibility
exceeds $99.5\,\%$. This indicates nearly perfect spatial overlap of the data
and program photons on PBS. The linear film polarizer placed before detector
$D_2$ is removed in further measurements and used in the state preparation
stage to ensure preparation of pure input state.

\begin{figure}[t]
\centerline{\includegraphics[width=0.85\linewidth]{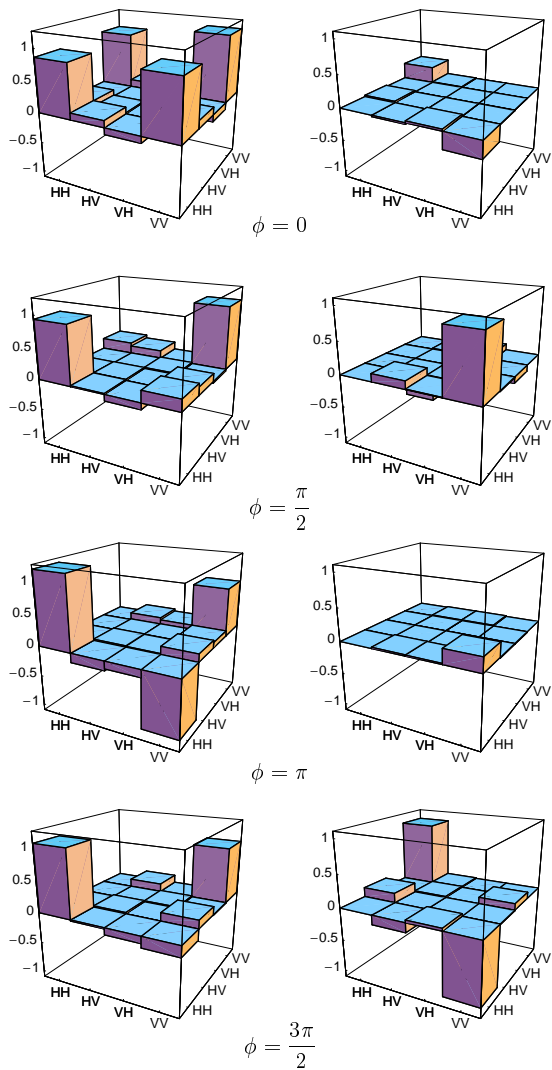}}
\caption{(Color online) Reconstructed process matrix $\chi$. Real (left column) and imaginary (right column)
parts of the reconstructed CP map $\chi$ are shown for four different values of the programmed 
phase shift $\phi$ encoded in the state of program photon.}
\label{gatefig3d}
\end{figure}

The success of the gate is heralded by a coincidence detection of a single
photon in each output port. In the experiment, we therefore measure the
coincidence rates $C_{13}$ between detectors D$_1$ and D$_3$ and $C_{23}$
between detectors D$_2$ and D$_3$. The width of the coincidence window is
set to 2~ns. For a given phase shift $\phi$ we characterize the performance
of the programmable gate by a tomographically complete measurement.
We set the state of the program photon to $\frac{1}{\sqrt{2}}(|H\rangle+e^{i\phi}|V\rangle).$
We then subsequently prepare the data photon in six different states
$|H\rangle$, $|V\rangle$, $|+\rangle$, $|-\rangle$, $|R\rangle$, and
$|L\rangle$,  where $|R\rangle=\frac{1}{\sqrt{2}}(|H\rangle+i|V\rangle)$
and $|L\rangle=\frac{1}{\sqrt{2}}(|H\rangle-i|V\rangle)$  denote the right
and left circularly polarized states, respectively. For each input we carry
out measurements for six different settings of the wave plates in the data
detection stage chosen such that the click of D$_1$ heralds projection
of the data photon onto the state $|H\rangle$, $|V\rangle$, $|+\rangle$,
$|-\rangle$, $|R\rangle$, and $|L\rangle$ by turns.
Every particular measurement takes 5~s and is repeated 10 times to gain
statistics. Average twofold coincidence rate is about 1300~s$^{-1}$.
The polarizer placed in front of the detector D$_1$ guarantees nearly ideal
projection onto a pure polarization state. Therefore, the presented results
were obtained using only $C_{13}$ coincidence data. Note that in this way
we do not need to precisely calibrate the relative detection efficiencies
of $D_1$ and $D_2$. We have confirmed that the results remain largely
unchanged if we use the coincidences $C_{23}$ instead or if we process
all data simultaneously. However, the results obtained from $C_{23}$ exhibit
slightly higher noise due to imperfect polarization filtering by the
polarizing beam splitter PBS$_\mathrm{M}$ \cite{PBSfootnote}.

\begin{figure}[!t!]
\centerline{\includegraphics[width=\linewidth]{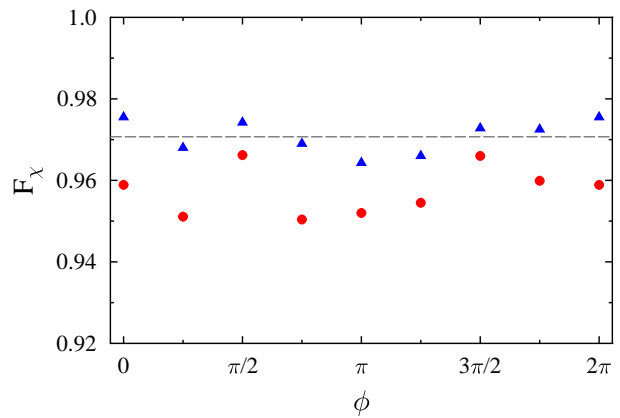}}
\caption{(Color online) Quantum process fidelity of the programmable gate is plotted as
a function of the encoded phase shift $\phi$. The fidelities before ($\bullet$)
and after ($\blacktriangle$) compensation of the constant phase offset
$\delta \phi$ are shown. The dashed line represents the best constant fit
to the compensated fidelity data with the value of 97{.}1\%.}
\label{processfidelityfig}
\end{figure}

From the experimental data we reconstruct the completely positive (CP) map
that fully characterizes the transformation of the data photon for a fixed
state of the program photon. We have performed the quantum process tomography 
for eight different phase shifts $\phi=\frac{k}{4}\pi$, $k=0,1,...,7$.
According to the Jamiolkowski-Choi isomorphism \cite{Jamiolkowski72,Choi75},
every CP map can be represented by a positive semidefinite operator $\chi$
on the tensor product of the input and output Hilbert spaces
$\mathcal{H}_{\mathrm{in}}$ and $\mathcal{H}_{\mathrm{out}}$. In our case
both $\mathcal{H}_{\mathrm{in}}$ and $\mathcal{H}_{\mathrm{out}}$ are
two-dimensional Hilbert spaces of polarization state of a single photon
hence $\chi$ is a $4\times 4$ matrix. The input state $\rho_{\mathrm{in}}$
transforms according to the formula $\rho_{\mathrm{out}}=\mathrm{Tr}_{\mathrm{in}}
[(\rho_{\mathrm{in}}^{T}\otimes \openone_{\mathrm{out}}) \, \chi]$,
where $T$ denotes transposition in a fixed basis. 
Due to the slight imperfections of the PBS, 
the implemented operation  is not exactly unitary and  may
involve some polarization filtering.
We therefore do not impose the constraint that $\chi$
has to be trace-preserving but allow for general trace-decreasing map
\cite{Cernoch06,Cernoch08}. We use the iterative maximum-likelihood
estimation algorithm that is described in detail elsewhere \cite{Jezek03,Paris04}.
This statistical reconstruction method yields a quantum process
$\chi$ that is most likely to produce the observed experimental data
\cite{Hradil97,Banaszek00}.

Figure~\ref{gatefig3d} displays the real and imaginary parts of the reconstructed
CP map $\chi$ for four different phase shifts $\phi=k \frac{\pi}{2}$, $k=0,1,2,3$.
We quantify the gate performance by  the process fidelity defined as follows,
\begin{equation}
F_{\chi}= \frac{\mathrm{Tr}[\chi \chi_{\mathrm{id}}(\phi)]}{\mathrm{Tr}[\chi]\mathrm{Tr}[\chi_{\mathrm{id}}(\phi)]}.
\label{Fidelitychi}
\end{equation}
Here $\chi_{\mathrm{id}}(\phi)$ is a process matrix representing the unitary
operation $U(\phi)$ (\ref{Uphi}),
\begin{equation}
\chi_{\mathrm{id}}(\phi)=\openone \otimes U(\phi) |\Phi^{+}\rangle \langle \Phi^{+}| \openone \otimes U^{\dagger}(\phi),
\end{equation}
where  $|\Phi^{+}\rangle=|H\rangle|H\rangle+|V\rangle|V\rangle$ denotes the
maximally entangled Bell state. Thus $\chi_{\mathrm{id}}$ is effectively
a density matrix of a pure maximally entangled state on
$\mathcal{H}_{\mathrm{in}}\otimes\mathcal{H}_{\mathrm{out}}$. The process
fidelity determined from the reconstructed CP maps is plotted in
Fig.~\ref{processfidelityfig} as a function of the phase shift $\phi$.
We can see that the fidelity is almost constant and exceeds $95\%$
for all values of $\phi$ which demonstrates very good functionality
of the programmable gate.

A careful analysis of the reconstructed CP maps reveals that the polarizing
beam splitter PBS imposes certain non-zero relative phase shift $\delta\phi$
between the vertical and horizontal polarizations. The active area of the PBS
where splitting of the vertical and horizontal polarization components occurs
is made of a stack of thin dielectric films. In principle, each of the polarization
and spatial modes passing through the PBS can acquire a different phase shift.
However, only a single effective combination of such phase shifts is relevant
in our experiment and gives rise to the phase offset $\delta\phi$.

\begin{figure}[!t!]
\centerline{\includegraphics[width=\linewidth]{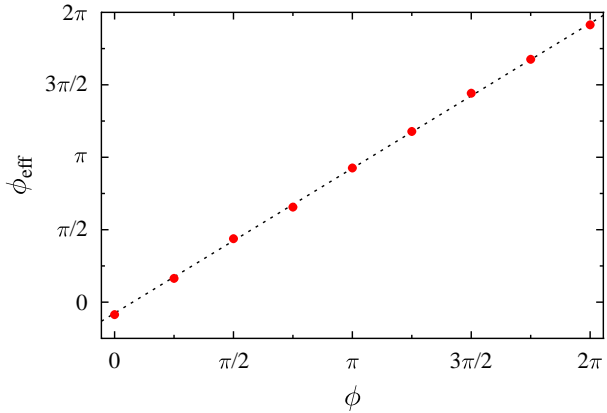}}
\caption{(Color online) Dependence of the effectively applied phase shift $\phi_{\mathrm{eff}}$
on the programmed phase shift $\phi$. The circles represent results obtained from
the reconstructed CP maps, the dashed line is the best linear fit to the data.}
\label{linearfit}
\end{figure}

We estimate the phase offset as follows. For each value of the encoded phase
shift $\phi$ we determine the effective applied phase shift $\phi_{\mathrm{eff}}$
by maximizing the overlap $\mathrm{Tr}[\chi 
\chi_{\mathrm{id}}(\phi_{\mathrm{eff}})]$ over $\phi_{\mathrm{eff}}$. 
The dependence of $\phi_{\mathrm{eff}}$ on $\phi$ is plotted in Fig.~\ref{linearfit}.
From the best linear fit to the data we obtain  $\delta\phi =-0.265\,\rm rad$.
This phase offset could be passively compensated e.g. by means of additional wave-plates
that would apply the relative phase shift $-\delta\phi$ to the output data photon.
We have carried out the software compensation and corrected the reconstructed CP maps
for the fixed phase offset. This calibration procedure increases the process fidelity
by about $1\%$ as shown in Fig.~\ref{processfidelityfig}. All the compensated fidelity
data are within one percent around the average value of $97.1\%$.
The achievable gate fidelity is mainly limited by the imperfections of the polarizing
beam splitter PBS that does not totally  reflect (transmit)  the V (H) polarization. 
The measured splitting ratios read $97.7:2.3$ and $0.5:99.5$ for vertical and horizontal
polarizations, respectively. A simple theoretical model predicts average process fidelity
$97.4\%$ which is in a very good agreement with the experimental results. 

Besides the quantum processes we have also reconstructed the single-qubit
output state for each input state. We have evaluated the state fidelity
$F=\langle\psi_{\mathrm{out}}|\rho|\psi_{\mathrm{out}}\rangle$ 
between the expected pure output state and the reconstructed (generally mixed)
state $\rho$. For each phase shift $\phi$ we average the state fidelity over
the six different input states to obtain the average state fidelity
$F_{\mathrm{avg}}$. We find that $F_{\mathrm{avg}}$ lies in the interval
$96.6\%-97.8\%$. The compensation of the phase offset $\delta\phi$ increases
the average state fidelity by almost $1\%$ to the range $97.6\%-98.5\%$. This
further confirms that the programmable gate operates with very high fidelity
for all values of the phase shift $\phi$ in the interval $[0,2\pi]$.

The average state fidelity $F_{\mathrm{avg}}$ and the process fidelity
$F_{\chi}$ exhibit almost perfect linear dependence of the form 
$F_{\mathrm{avg}}=0.727F_{\chi}+0.275$. This is consistent with the
theoretically predicted relation between these two fidelities for
deterministic processes \cite{Horodecki99},
$F_{\mathrm{avg}}=\frac{1}{3}(2 F_{\chi}+1)$. The observed discrepancy
is mainly due to the fact that we perform independent maximum likelihood
reconstructions of the quantum process and output states
while the theoretical formula assumes that the output states are calculated
from the input states using the process matrix $\chi$.
Also, the reconstructed CP map is not exactly trace preserving.

\begin{figure}[!t!]
\centerline{\includegraphics[width=\linewidth]{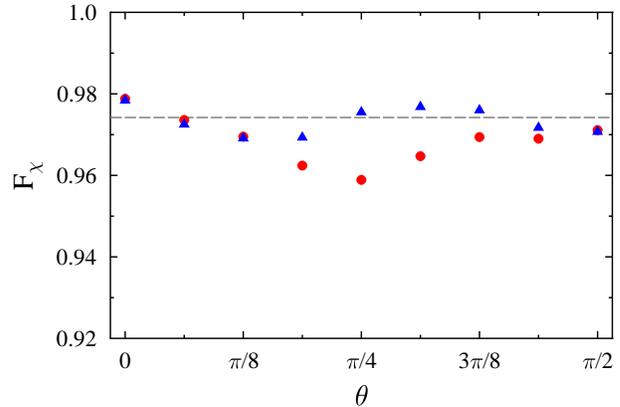}}
\caption{Quantum process fidelity of the programmable partial polarization filter
is plotted as a function of the filtering angle $\theta$. The fidelities before
($\bullet$) and after ($\blacktriangle$) compensation of the constant phase offset
$\delta \phi$ are shown. The dashed line represents the best constant fit to the
compensated fidelity data with the value of $97.4\%$.}
\label{filterfig}
\end{figure}

We next show that our device can also function as a programmable partial
polarization filter \cite{Hillery04}. For this purpose we prepare the program qubit in various
linear polarization states $\cos\theta|H\rangle+\sin\theta|V\rangle$. Repeating
the calculation leading to Eq.~(\ref{psiout}) we find the (non-normalized)
output state of the data qubit to be 
\begin{equation}
|\psi_{\mathrm{out}}\rangle_D=\alpha\cos\theta|H\rangle+\beta\sin\theta|V\rangle.
\end{equation}
The amplitude of vertical polarization is attenuated (or amplified) by a factor
of $\tan\theta$ with respect to amplitude of the horizontal polarization.
We carry out the complete quantum process tomography of the programmable quantum
filter for nine different values of $\theta=\frac{n}{16}\pi$, $n=0,1,...,8$.
The process fidelity can be calculated according to Eq. (\ref{Fidelitychi})
where the ideal filtering operation is now described by a partially entangled state,
$\chi_{\mathrm{id,filter}}(\theta)=(\cos\theta|H\rangle|H\rangle+\sin\theta|V\rangle|V\rangle)
(\langle H|\langle H|\cos\theta+\langle V|\langle V| \sin\theta)$. The experimentally
determined process fidelity is plotted in Fig.~\ref{filterfig}. Similarly as for
the programmable unitary gate, the compensation of the constant phase offset $\delta\phi$
increases the fidelity. The improvement is most significant for $\theta=\pi/4$ while for
complete filtering ($\theta=0$ and $\theta=\pi/2$) the phase shift is irrelevant and its
compensation does not change the fidelity.

In conclusion, the programmable single-qubit phase gate working on single-photon
polarization-encoded qubits has been proposed and experimentally
developed. The gate operation has been thoroughly tested by complete quantum process
tomography. The comparison of the reconstructed processes and the corresponding 
theoretical ones yields high process fidelity of about 97\% with negligible dependence 
on the encoded phase shift. It has been demonstrated that with 
a different set of program states, the device can also
operate as a programmable partial polarization filter.
The implemented programmable gates can serve as building blocks of more
complex multi-qubit linear-optics quantum gates or other optical quantum information processing devices.

\begin{acknowledgments}
We would like to thank to Lucie Bart\r{u}\v{s}kov\'{a} 
for help and fruitful discussions during the experiment. This work has
been supported by Research Projects  ``Center of Modern
Optics'' (LC06007) and ``Measurement and Information in Optics''
(MSM 6198959213) of the Czech Ministry of Education.
\end{acknowledgments}


\begin{thebibliography}{99}
\bibitem{PBSfootnote}
The  process and state fidelities determined using coincidences $C_{23}$ are about $1\%$
lower than the fidelities obtained from coincidences $C_{13}$.
\end{thebibliography}
\end{document}